\documentclass[]{aa}
\usepackage{graphicx}
\usepackage{epsfig}
\usepackage{txfonts}
\def\lae{\mathrel{<\kern-1.0em\lower0.9ex\hbox{$\sim$}}}
\def\gae{\mathrel{>\kern-1.0em\lower0.9ex\hbox{$\sim$}}}
\begin{document}
   \title{An X-ray bright ERO hosting a type~2 QSO\thanks{Based on data 
   obtained with the European Southern Observatory Very Large
   Telescope, Paranal, Chile, program 075.B-0229(A).}}

   \author{P. Severgnini\inst{1}
          \and
	  A. Caccianiga\inst{1} 
	  \and
	  V. Braito\inst{1,2}
	  \and
	  R. Della Ceca\inst{1} 
	  \and
	  T. Maccacaro\inst{1}
	  \and
	  M. Akiyama\inst{3}
	  \and
	  F. J. Carrera\inst{4}
	  \and 
	  M. T. Ceballos\inst{4}
	  \and
	  M. J.  Page\inst{5}
	  \and
	  P. Saracco\inst{1}
	  \and
	  M. G. Watson\inst{6}
	   }

   \offprints{P. Severgnini, e-mail: paola@brera.mi.astro.it}

   \institute{INAF - Osservatorio Astronomico di Brera (OAB),
              Via Brera 28, I-20121, Milano, Italy\\
	      \email{paola, caccia, braito, rdc, tommaso, saracco@brera.mi.astro.it}
	   \and
	      Exploration of the Universe Division, NASA Goddard Space Flight
              Center, Greenbelt Road, Greenbelt, MD 20771 \\
	      \email{vale@milkyway.gsfc.nasa.gov}
	  \and
	      Subaru Telescope, National Astronomical Observatory of Japan,
	      650 North A'ohoku Place, Hilo, HI 96720, USA\\
	      \email{akiyama@subaru.naoj.org}
	  \and
	      Instituto de Fisica de Cantabria (CSIC-UC), Avenida de los 
	      Castros, 39005 Santander, Spain\\
	      \email{carreraf, ceballos@ifca.unican.es}
	  \and
	      Mullard Space Science Laboratory, University College London, 
	      Holmbury St. Mary, Dorking, Surrey RH5 6NT, UK\\
	      \email{mjp@mssl.ucl.ac.uk}
	  \and
	      X-ray Astronomy Group, Department of Physics and Astronomy, 
	      Leicester University, Leicester LE1 7RH, UK\\
	      \email{mgw@star.le.ac.uk}
	      }

   \date{Received ...; accepted ...}

   \abstract{We present the XMM-Newton and the optical--VLT spectra along with
the optical and the near--infrared photometric data of one of the brightest
X-ray  (F$_{2-10\rm keV}$$\sim$10$^{-13}$ erg s$^{-1}$ cm$^{-2}$) extremely red
objects (R-K$\ge$5)  discovered so far. The source, XBS J0216-0435, belongs to
the  {\it XMM-Newton Bright Serendipitous Survey} and it has extreme
X--ray--to--optical ($\sim$220) and X--ray--to--near--infrared ($\sim$60) flux
ratios. Thanks to its brightness, the X--ray statistics are good enough for an
accurate spectral analysis by which the presence of an X--ray obscured
(N$_H$$>$10$^{22}$ cm$^{-2}$) QSO (L$_{2-10\rm keV}$=4$\times$10$^{45}$ erg
s$^{-1}$) is determined. A statistically significant ($\sim$99\%)  excess
around 2~keV in the observed--frame suggests the presence of an emission line.
By assuming that this feature corresponds to the iron K$\alpha$ line at
6.4~keV, a first estimate of the redshift of the source is
derived (z$_X$$\sim$2). The presence of a high redshift QSO2 has been finally  confirmed
through dedicated   VLT optical spectroscopic observations
(z$_O$=1.985$\pm$0.002). This result yields to an optical validation  of a new
X--ray Line Emitting Object (XLEO) for which the redshift has
been firstly derived  from the X--ray data. XBS J0216-0435 can be
considered  one of the few examples of X--ray obscured QSO2 at high redshift
for which a detailed X--ray and optical spectral analysis has been possible.
The spectral energy distribution from radio to X--rays is also presented. 
Finally from the near--infrared data the luminosity and the stellar mass of the
host galaxy has been estimated finding a new example of the coexistence at
high--z between massive galaxies and powerful QSOs.

    }

   \titlerunning{X-ray bright ERO with an extreme X/O
   flux ratio}

   \maketitle
%
%________________________________________________________________

\section{Introduction} 

\begin{figure*}[th!]
\begin{center}
\begin{tabular}{cc}
\epsfig{file=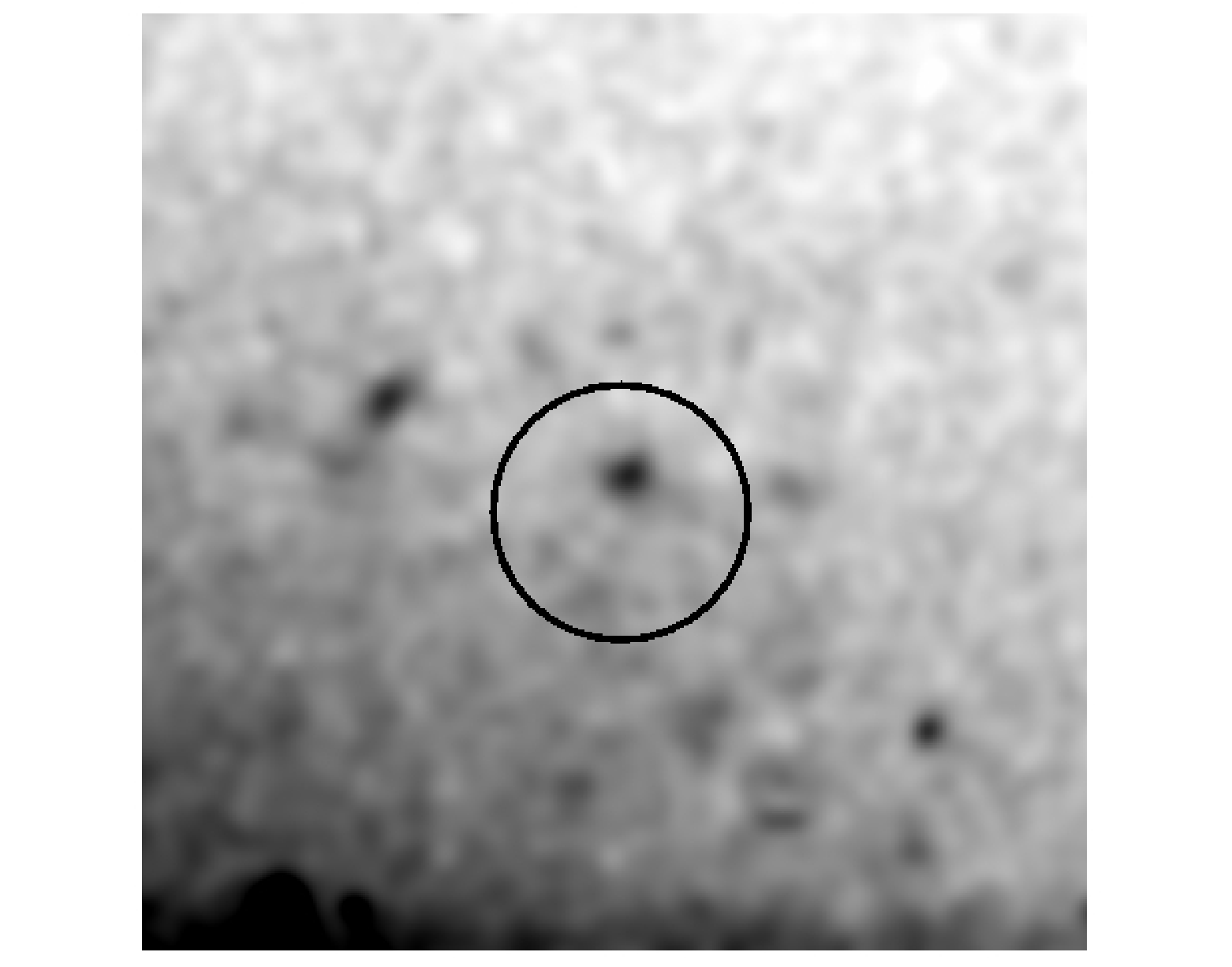,width=0.3\textwidth}&\hskip +0.5truecm
\epsfig{file=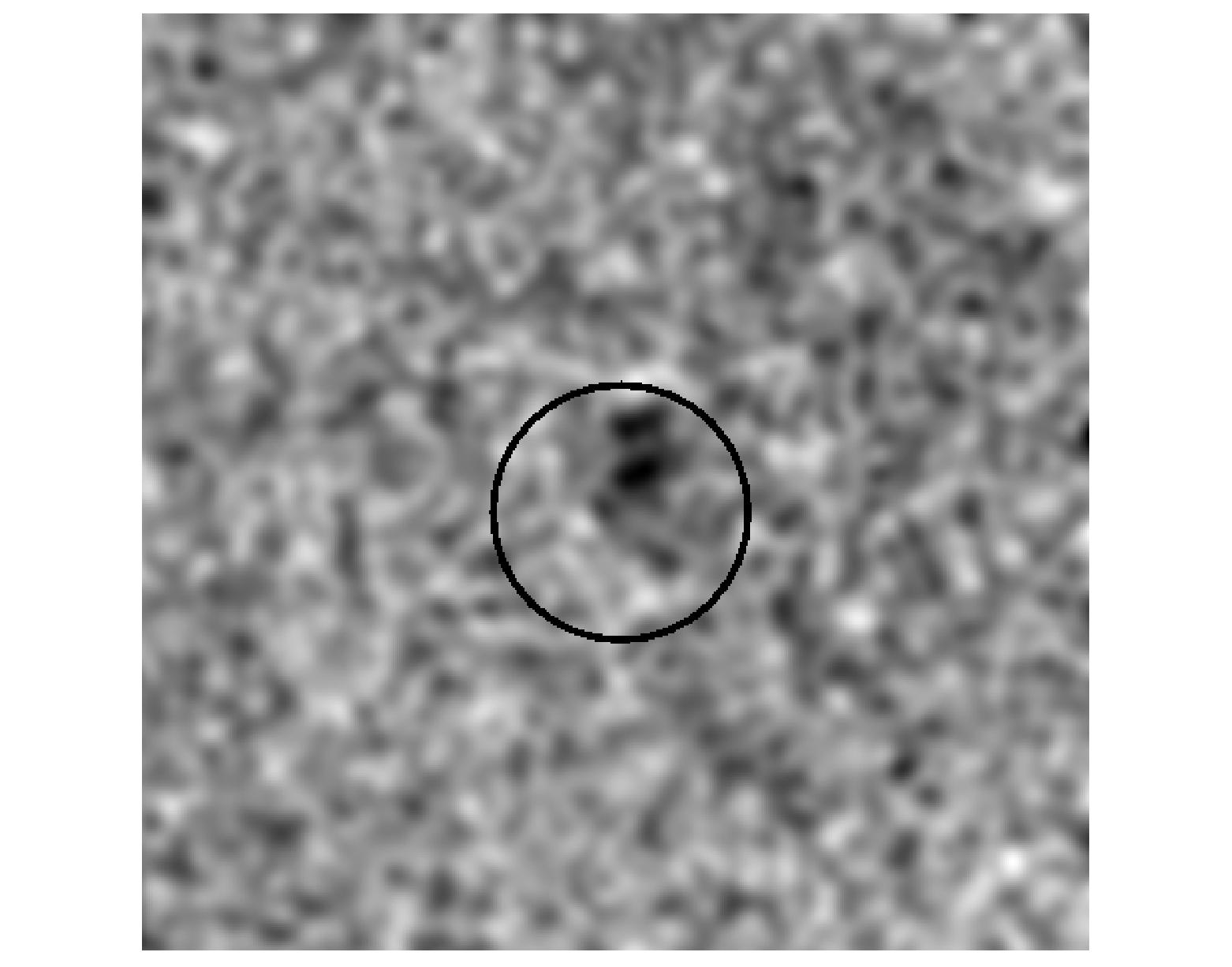,width=0.295\textwidth}\\
\end{tabular}
\end{center}

\caption{R-band (left panel) and K-band (right panel) smoothed images 
(30$^{\prime\prime}$$\times$30$^{\prime\prime}$) of XBS J0216-0435. North is to
the top  and east to the left. The circle of 4$^{\prime\prime}$ of radius marks
the centroid of the X--ray position. The strong gradient present in the optical
image is due to a very bright star (R=9 mag) near the objects considered here.}
\end{figure*}

Extremely red objects (R-K$\ge$5, EROs) with bright X--ray fluxes
(F$_{(2-10~keV)}$$\ge$10$^{-14}$ erg s$^{-1}$ cm$^{-2}$) and with  high
X--ray--to--optical and X--ray--to--NIR (near--infrared) flux  ratios
(F$_{(2-10~keV)}$/F$_{opt}$, F$_{(2-10~keV)}$/F$_{K}$$>>$10) are among the best
candidates to be  obscured\footnote{We use here N$_H$=10$^{22}$ cm$^{-2}$ and
Av$\ge$1.7 mag as dividing values between unabsorbed and absorbed sources in
the X--ray and optical bands respectively.}  AGNs, i.e. the main ingredient of
the X--ray Cosmic Background (Gilli et al. \cite{gilli}; Ueda et al.
\cite{ueda}). Indeed, given the  minimum  redshift observed for extragalactic
EROs with such high flux ratios (z$\gae$0.6, see e.g.  Akiyama et al.
\cite{akiyama01};  Mignoli et al. \cite{mignoli}; Severgnini et al.
\cite{severgnini05a} and references therein; Maiolino et al.
\cite{maiolino06})  an X--ray flux $\sim$10$^{-14}$ erg s$^{-1}$ cm$^{-2}$
corresponds to  L$_{2-10\rm keV} >$10$^{43}$ erg s$^{-1}$, typical of AGN. 
Moreover, a factor of at least 10 higher in the  F$_{(2-10~keV)}$/F$_{opt}$ and
F$_{(2-10~keV)}$/F$_{K}$ ratios than those observed for  unobscured AGNs
(F$_{(2-10~keV)}$/F$_{opt}$, F$_{(2-10~keV)}$/F$_{K} \sim$1) can be easily
justified by invoking the presence of  an obscuring optically--thick medium
(e.g. molecular torus) along the line of sight. In this case, while the UV and
optical emission is totally or heavily suppressed by the large amount of dust,
the X--ray emission is less affected by the absorbing medium producing high
F$_{(2-10~keV)}$/F$_{opt}$ and  F$_{(2-10~keV)}$/F$_K$. If the above scenario
is true, it implies that the selection of  EROs with such high flux ratios at
bright X--ray fluxes (F$_{(2-10~keV)}$$\ge$10$^{-13}$ erg s$^{-1}$ cm$^{-2}$) 
is an extremely efficient way to find  the long sought  high luminosity
(L$_{(2-10~keV)}$$>$10$^{44}$ erg s$^{-1}$), X-ray obscured type 2 
QSOs\footnote{The "type 2"  denomination refers to the optical spectral
properties of the AGNs  characterized only by strong and highly ionized narrow 
(FWHM$<$1000 km/s) emission lines.} (hereafter QSO2).

In this paper we present XMM-Newton and VLT optical spectroscopic data  of a
very red X--ray selected source (XBS J0216-0435) belonging to  the {\it
XMM-Newton Bright Serendipitous Survey} (XMM--BSS, Della Ceca et al.
\cite{dellaceca05}; Della Ceca et al.
\cite{dellaceca04}; Caccianiga et al. \cite{caccianiga};  Severgnini et al.
\cite{severgnini03}). This source, given its  bright X-ray flux and its high
F$_{(2-10~keV)}$/F$_{opt}$ and F$_{(2-10~keV)}$/F$_{K}$, represents  the
prototype of the EROs in which a QSO2 is expected.  The good X--ray statistics
have made the spectral analysis possible  and allowed us to give a first
estimate  of the redshift of the source (z$\sim$2, Sect.~3). The X--ray properties
found  (the spectral shape, the amount  of absorption and the presence of the
FeK line) are typical of an obscured QSO. The presence of a high redshift 
QSO2 has been finally  confirmed through dedicated   VLT optical spectroscopic
observations (Sect.~4).

Throughout this paper we assume H$_0$=65 km s$^{-1}$ Mpc$^{-1}$ and
$\Omega_M$ =0 3, $\Omega_{\Lambda}$=0 7. All the magnitudes are in the Vega
system.

%__________________________________________________________________

\section{An ERO with extreme F$_{(2-10~keV)}$/F$_{opt}$ and 
F$_{(2-10~keV)}$/F$_{K}$ flux ratios}

XBS J0216-0435 belongs to the XMM-BSS sample. The sample definition and
selection criteria are described in Della Ceca et al. (\cite{dellaceca04}). The
X--ray source discussed here is in the  XMM-Newton Subaru field 
SXDS4\footnote{No  Subaru data are available since the object is
outside the Suprime-cam deep survey field.} and has a  count rate in the 
0.5--4.5 keV band of (1.32$\pm$0.12)$\times$10$^{-2}$ counts/s which corresponds to a
F$_{(0.5-4.5~keV)}\sim$10$^{-13}$ erg cm$^{-2}$ s$^{-1}$ (see Della
Ceca et al. \cite{dellaceca04}). Contrary to the majority ($\sim$90\%) of the
XMM-BSS sources, the optical counterpart of XBS J0216-0435 is much 
fainter than the POSS~II limit (R$\sim$21 mag).

Our own R--band and K--band photometric observations have been performed  on
January 2005 at the ESO New Technology Telescope (NTT) using the ESO Multi-Mode
Instrument (EMMI, 1 hour of exposure time) and on October 2002 at the
Telescopio Nazionale Galileo (TNG) using the Near Infrared Camera Spectrometer
(NICS, 15 minutes of exposure) respectively.  Both the observing runs 
were
carried out under similar seeing conditions ($\sim$1.2$^{\prime\prime}$). The
optical and NIR images were reduced using standard {\it iraf}
routines. The zero-points have been derived by measuring the  instrumental
magnitudes of standard stars observed just before and/or after  the scientific
target and assuming the average extinction reported in the Observatory web
pages. Astrometric calibration was performed using the software package GAIA
(version 2.6--9 by P.W. Draper) by matching sources found in the USNO catalogue
(Monet et al. \cite{monet}). The typical uncertainty in the best fit solution
is about 1$^{\prime\prime}$. A weak (R$\sim$24.5 mag, see Figure~1, left panel)
optical counterpart is  visible within 4 arcsec   from the X--ray position of
the source (corresponding to the 90\% confidence level X--ray error circle, 
see Della Ceca et al. \cite{dellaceca04}).  Within the same distance from the
X--ray position, two possible weak NIR objects are visible with  a
K'$\sim$19.5 mag (see Figure~1, right panel).   The two NIR sources
are separated by $\sim$1$^{\prime\prime}$.   Even though 
the quality of the data is
not good enough to allow a morphological analysis, all the sources appear
extended. We have positionally registered the R to the K' image by using the 8
brightest sources in the field. We find that the optical source is spatially
coincident with the southern NIR source. This latter is also the
NIR source closest to the X--ray centroid position (less than
1$^{\prime\prime}$).  Given the  color of the two sources (R-K'$\sim$5 for the
southern source and R-K'$>$6 for the northern one) these objects  can be
classified as EROs. Although the different optical--to--NIR colors
of the two objects suggest that they could have different redshifts,  the
presence of a group of galaxies can not be ruled out. However, even in the case
of a group, the X--ray emission observed is  dominated by a powerful AGN (see
Sect.~3) which, on the basis of VLT optical spectroscopic observations (see
Sect.~4), is hosted in the southern ERO. 
\begin{figure}[th!]
{\centerline{\epsfig{file=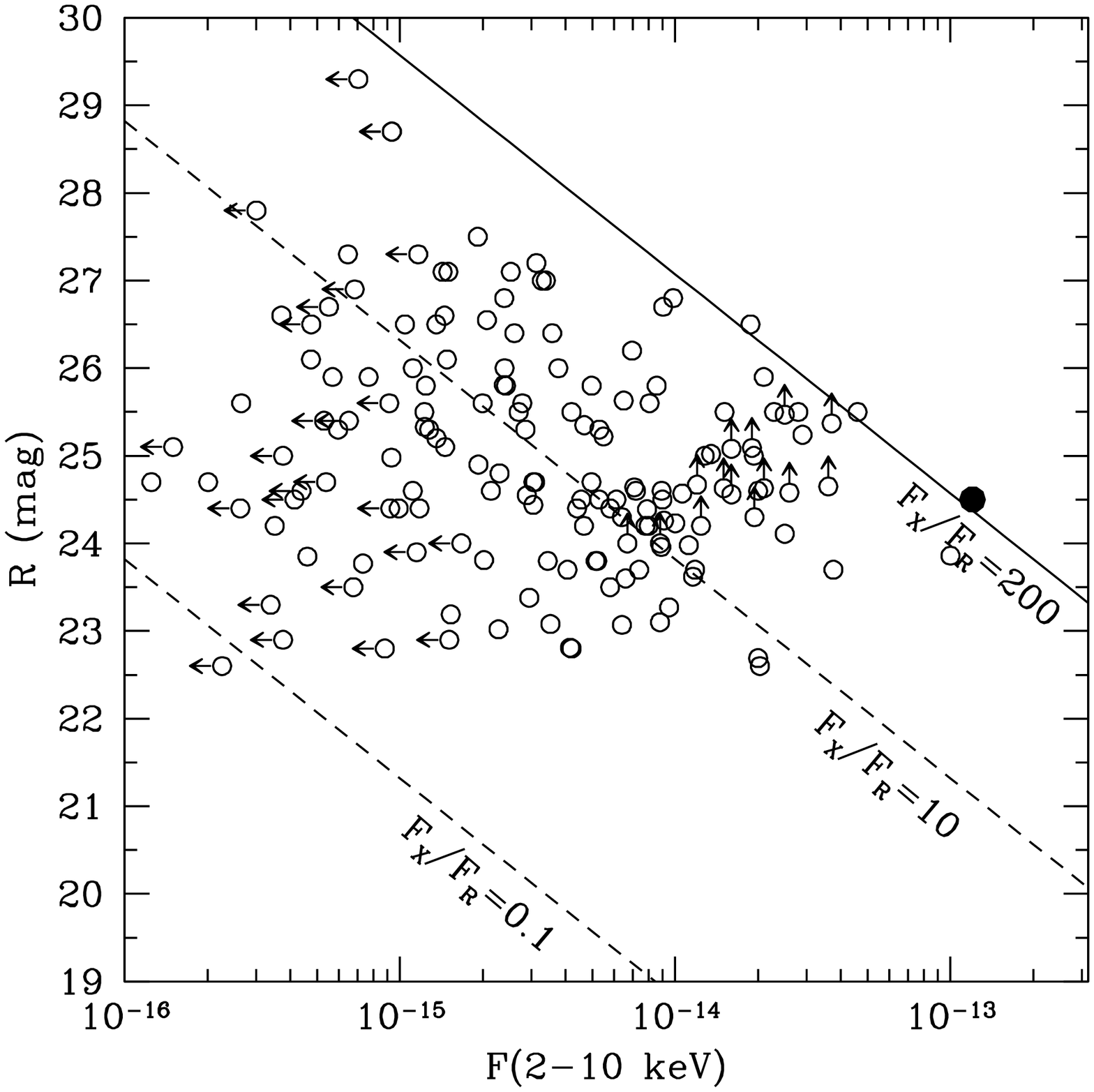,width=0.5\textwidth}}}

\caption{R magnitude vs. 2--10 keV flux for XBS J0216-0435 (filled
circle)  and for other X-ray emitting EROs (empty circles)  taken from the
literature (Mignoli et al. \cite{mignoli}; Mainieri et al.
\cite{mainieri}; Stevens et al. \cite{stevens};  Alexander et al.
\cite{alexander02}, \cite{alexander03}; Barger et al.\cite{barger03}; Roche et
al. \cite{roche}; Szokoly et al. \cite{szokoly}; Willott et al. \cite{willott};
Gandhi et al. \cite{gandhi}; Severgnini et al. \cite{severgnini05a}, Brusa et
al. \cite{brusa05}). Upper and lower limits of the (2--10 keV) and R magnitudes
are marked with arrows.  The two dashed lines define the
region where unobscured type~1 AGNs typically lie (see e.g. Maccacaro et al.
\cite{maccacaro88};  Fiore et al. \cite{fiore}).}

\label{fig3}
\end{figure}
This source is the object with the 
highest F$_{(2-10~keV)}$/F$_{opt}$\footnote{F$_{(2-10~keV)}$ has been derived
from the F$_{(0.5-4.5~keV)}$ by assuming an observed $\Gamma$=1.5 on the basis
of the X--ray colors, see Della Ceca et al. (\cite{dellaceca04}). F$_{opt}$ is the
integrated flux in the Cousin R band system. More specifically:
F$_{opt}=1568~\AA~\times (2.15\times 10^{-9}$ cgs$~\AA^{-1} \times 10^{(-0.4
\cdot m_R)}$), see Fukugita et al. \cite{fukugita}.} ($\sim$220) among the
XMM--BSS sources (see Fig.~6 in Della Ceca et al. \cite{dellaceca04}). 
Furthermore, taking into account the  2--10 keV X--ray
emitting EROs collected from the literature by  Brusa et al. (\cite{brusa05}), 
XBS J0216-0435 is one of the EROs with  the highest X--ray flux among those
with  the highest F$_{(2-10~keV)}$/F$_{opt}$ (filled circle in Figure~2)  and
X--ray--to--NIR ratios (XBS J0216-0435 has a
F$_{(2-10~keV)}$/F$_K$\footnote{For the K-band we adopted the   magnitude to
flux conversion used in Brusa et al. \cite{brusa05}.}$\simeq$60,  see for
comparison Fig.~4 of Brusa et al. \cite{brusa05}). We note that, as described
in the next section, we have derived the X--ray flux from two different {\it
XMM-Newton} observations separated by more than two years. The extremely high
F$_{(2-10~keV)}$/F$_{opt}$(F$_K$)  ratios quoted above are not likely due to an
intrinsic variation of the source since the X--ray fluxes measured in the 
two observations are consistent within the errors.

%______________________________________________________________
\section{XMM-Newton data}
\subsection{X-ray observations}

XBS J0216-0435 was serendipitously observed by the EPIC cameras on 08 August
2000 (OBS ID=0112371701, $\sim$24 ksec exposure) and on 07 January 2003 (OBS
ID=0112372001, $\sim$28 ksec exposure).  The observations were carried out  in
full frame mode and with the thin filter applied. The data have been processed
using the SAS  ({\it Science Analysis Software}) version 5.4.  The latest
calibration files and response matrices released by the EPIC  team have been
used to create new response matrices that include also the correction for the
effective area of the source position in the detector (off--axis $\sim$8
arcmin). Events files released from the standard  pipeline have been  filtered
from high background time intervals and only events corresponding  to patterns
0--12 and 0--4 have been used for the MOS and pn respectively. In the first of
the two exposures, XBS J0216-0435 lies on a gap of the pn camera and only the
MOS data are available.  The net exposure times after data cleaning are
reported in Table~1.

At the spatial resolution of XMM-Newton EPIC instruments, our source is
consistent with a point--like object.  This result does not discard the
presence of a group of galaxies or a small cluster (core radius of order of
90 kpc, Mohr et al. \cite{mohr}).

\begin{table}[t!]
\begin{center}
\caption{XMM-Newton observations of XBS J0216-0435}
\label{obs_log}
\small
\begin{tabular}{cccc}
\hline \hline   
Obs. ID & EPIC-cameras & Net exposure time \\
 011237                &             &  [ksec]  \\
\hline
\hline 
1701       & MOS1+MOS2 &  $\sim$39  \\
		 &	   	\\
\hline 
2001       &    pn     &  $\sim$22  \\ 
	         & MOS1+MOS2 &  $\sim$52   \\
\hline 
\hline
\end{tabular}
\end{center}
\end{table}

\subsection{X-ray spectral analysis}

The spectral analysis has been performed using XSPEC 11.2.0. The X-ray
spectrum was extracted using a circular region  of 28$^{\prime\prime}$ of
radius in all the images except for the MOS data of the second observation
where a smaller aperture (21$^{\prime\prime}$ of radius)  has been used
because of the proximity of a CCD gap. In all cases the background spectra
were extracted from $\sim$4 times larger source-free circular regions
close to the object. In order to improve the statistics, MOS1 and MOS2 data of
each observation have been combined together and, finally,  MOS and pn spectra
have been binned to have at least 20 total counts per energy channel. The MOS
and pn spectra of the second observation  were fitted simultaneously in
the 0.5--10 keV band leaving free the relative normalizations. Total net
counts of $\sim$250 and $\sim$740 have been accumulated in the first and
second observation respectively. In the fitting procedure, the appropriate
Galactic hydrogen column density along the line of sight 
(2.37$\times$10$^{20}$ cm$^{-2}$) has been taken into account (Dickey \&
Lockman \cite{dickey}). In the rest of the paper, and unless  stated
otherwise, errors are given at the 90\% confidence level for one interesting
parameter ($\Delta \chi$$^2$=2.71).

The photometric optical, NIR and X--ray properties of XBS J0216-0435  suggest
that the observed X--ray emission  could be due to the presence of an obscured
AGN  and/or a group or a small cluster of galaxies. While  a pure thermal
component (mekal thermal plasma model, Mewe, Gronenschild \& van den Oord
\cite{mew}, Kaastra \cite{kaa}, Liedahl, Osterheld \& Goldstein \cite{lie}) is
rejected by the fit (assuming solar abundances and kT$<$10 keV the resulting
$\chi^2/dof$ is higher than 2.5 for any z), by fitting the data with a single
absorbed  power--law model (typical of obscured AGN)  we find a good
description of the overall spectrum of the source (for both the observations
we obtained a $\chi^2/dof \sim$1).  The addition of a thermal  component to
the power-law model is not statistically required.  These results  make
it unlikely that the X--ray emission is associated with a cluster/group of
galaxies, and point towards the presence of a powerful, dominant AGN.  
Furthermore, as already discussed in Severgnini et al. \cite{severgnini05b}
and  shown in Figure~3, by fitting the data   with an absorbed power--law
model, the presence of an excess around 2~keV (observed--frame) suggests
line-like residuals.  It is
worth noting that the line--like residual has been observed only in the second
observation where the statistics are better. We have verified that  the
equivalent width estimated for this line (see Table~2) is comparable to the
upper limit measurable in the first observation. The data of the second
observation have been fitted again adding a Gaussian component ($\Delta
\chi^2$=10 for $\Delta$ d.o.f=3).  Since the strongest feature expected in the
X--ray spectrum of AGN is the neutral iron K$\alpha$ line at 6.4~keV
rest--frame  (Reynolds et al. \cite{reynolds}), we have associated the 2~keV
excess with this line. This places the source at z=2.27$^{+0.11}_{-0.35}$. 

The optical VLT spectrum obtained  for the object visible in the R-band
(Fig.~1, left panel) gives a redshift of z=1.985 (see Sect.~4). This value is
in excellent  agreement with the one derived from the XMM-Newton data. We
report in Table~2  the best fit (rest--frame) X--ray parameters and the
intrinsic X--ray  luminosity obtained by fixing the redshift to 1.985.  These
values refer to the data of the  second observation where the line is
observed. Although the quality of the data  from the 
first observation are not good enough
to detect the Fe line (we don't use them in this part of the
analysis), the continuum parameters obtained from the two observations are
consistent within the errors.

On the basis of the optical redshift,  the rest--frame energy of the line
falls to  5.8$^{+0.7}_{-0.2}$ keV. This value is consistent with the
rest--frame energy of the iron K$\alpha$ line (6.4 keV)\footnote{Two other  excesses of lower statistical
significance ($\Delta \chi^2$=4) are present at $\sim$2.37 keV and 3.19 keV
(7.07 keV and 9.52 keV in the rest--frame assuming a z=1.985). The first one
could be associated to the ionized iron K$\beta$ line (7.05 keV)}.  A detailed analysis
of the line properties is not warranted by the quality of the data. Both  the
line broadening and the high EW$_{Fe-k \alpha}$  observed (see Table~2) could
be due to the  merging of  several narrow ($\sigma$=0.1 keV)  lines (e.g. 6.4
keV plus 6.7 keV).  As reported in Table~2, the intrinsic column density
measured is typical of obscured AGN (N$_H$$\ge$10$^{22}$ cm$^{-2}$) and the
intrinsic luminosity places the source in the QSO regime (L$_{(2-10
keV)}$=4.1$\times$10$^{45}$ erg s$^{-1}$).

In conclusion, the X--ray spectral analysis unambiguously reveals the
presence of a dominant, high redshift and X--ray obscured QSO.

\begin{figure}[t!!!]
{\centerline{\epsfig{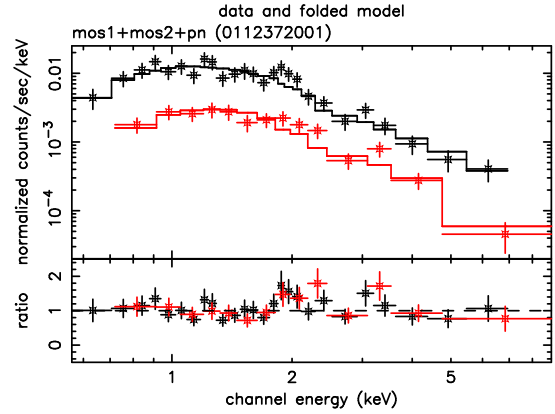}}}
\caption{Mos+pn folded X--ray spectrum of XBS J0216-0435 (upper panel, solid
points) obtained during the second (0112372001) observation. The data are 
fitted by a single absorbed power-law model (continuous lines).  
Ratio between data and model is plotted in the lower panel as
a function of energy.}
\label{fig3}
\end{figure}

\begin{table*}[t!!!!!] 
\caption{Rest--frame best fit parameters  assuming an absorbed power--law at 
z=1.985.}
\label{obs_log} 
\small
%\small
\begin{tabular}{ccccccccccccccccccccccc} 
\hline
Obs. ID & $\Gamma$ & $N_{\rm H}$ & E$_{Fe-k \alpha}$ & EW$_{Fe-k \alpha}$ &
$\sigma_{Fe-k \alpha}$ & $\chi^2/dof$ & F$_{2-10\rm keV}$$^a$$^c$&
L$_{2-10\rm keV}$$^b$$^c$ \\ 
011237 &          & $[10^{22}cm^{-2}]$ & [keV] & [keV] &  [keV]&
&[10$^{-13}$ erg cm$^{-2}$ s$^{-1}$] & [10$^{44}$ erg s$^{-1}$] \\

\hline
\hline
~~\\
 2001  & 2.0$^{+0.2}_{-0.1}$ &   4.7$^{+1.5}_{-1.3}$  & 5.8$^{+0.7}_{-0.2}$ &
0.7$^{+0.5}_{-0.4}$ & 0.3$^{+0.7}_{-0.3}$ &
25.11/33 & 1.1 & 41.0\\
~~\\
\hline
\end{tabular}
~~~\\
$^a$ The flux is corrected only for the Galactic absorption.\\
$^b$ The luminosity is corrected for Galactic and intrinsic absorption.\\
$^c$ Statistical uncertainties are of about 20\%.\\
\end{table*}

\begin{figure}[t!]
{\centerline{\epsfig{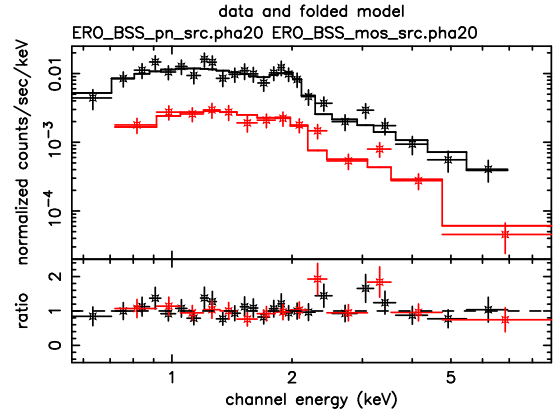}}}

\caption{Mos+pn folded X--ray spectrum of XBS J0216-0435 (upper panel, solid
points) and the best--fit for the following model: single absorbed power-law
model plus Gaussian line at the optical redshift of z=1.985 (continuous lines).
Ratio between data and model is plotted in the lower panel as a function of
energy.}
\label{fig3}
\end{figure}

\section{Optical spectroscopic data}
\subsection{Observation and data reduction}

The object visible in the R-band (Fig.~1, left panel)  has been observed
spectroscopically at ESO VLT using FORS2 during  the night between the 29th and
the 30th of September 2005 (from UT=04:14:12 to UT=05:14:38, including 
overheads). Grism 300V+20 was used with no sorting filters and a slit width of
1.6$\arcsec$. The  grism gives a spectral dispersion of 3.3 A pixel$^{-1}$. The
slit was rotated -20 $\deg$ in order to include the second object possibly
observed in the K-band image. However, only one source turned out to be visible
in the spectrum. Two exposures of 2790 seconds each have been taken for a total
of 5580 seconds. The DIMM seeing  during the observation ranges from
0.90$\arcsec$ to 1.58$\arcsec$ with an average value of about 1.2$\arcsec$. The
slit was not oriented along the parallactic angle, but the  source was observed
at a low airmass (between 1.1 and 1.25) and the slit width was large
(1.6$\arcsec$) when compared to the average seeing so that only a marginal flux
loss in the blue part of the spectrum may have occurred. Therefore, no
correction has been applied. For the data reduction we have used the standard
IRAF {\it long-slit} package following the standard steps.

\subsection{Spectral analysis}

The flux-calibrated spectrum is shown in Fig~5. At least three emission lines
are clearly visible in the spectrum at 4624\AA, 4899\AA\  and 5696\AA. These 3
lines are identified with C IV at $\lambda_{rest}$=1549\AA\ (z=1.985), He II
at 1640\AA\ (z=1.987) and C III] at 1909\AA\ (z=1.984) respectively. Two
weaker additional lines are then possibly identified at $\sim$3960\AA\ with C
II at $\lambda_{rest}$=1336\AA\ and at $\sim$8350\AA\ with Mg II at 
$\lambda_{rest}$=2798\AA. Using the 3 strongest lines we estimate a redshift of
z=1.985$\pm$0.002, in excellent agreement with the one derived from X--rays.

The computed widths for the three strongest lines are between 900 km s$^{-1}$
(He II) and  1400-1500 km s$^{-1}$ (CIV and CIII]). Considering the spectral
resolution of the instrument the intrinsic width of the lines is consistent
with being $\leq$1200 km s$^{-1}$. This is close to the ``standard''
classification of type 2 AGN (FWHM$<$1000 km s$^{-1}$) while it is different
from the line widths measured in ``Narrow Lined QSO'' (NLQSO) like those
studied by Baldwin et al. (\cite{baldwin88}) (1700 - 2700 km s$^{-1}$).
Another strong indication of the fact that XBS J0216-0435 is a ``truly'' type
2 QSO and not a QSO where the broad line region produces intrinsically narrow
lines, is given by the He II/CIV line ratio. As suggested by Heckman et al.
(\cite{heckman95}) the He II/CIV line ratio can be used to distinguish between
true type 2 objects and narrow line QSO. The former sources show a very
strong He II (He II/CIV $\sim$0.6-0.9) while in the latters the HeII/CIV ratio
is much lower ($\sim$0.05 - 0.2). The optical spectrum of XBS J0216-0435 shows
a very prominent He II line (He II/CIV $\sim$0.6) in full agreement with the
values observed in type 2 AGN (type 2 QSO, Sy2 and Narrow Line radio
galaxies).

We conclude that, from the optical spectral point of view, XBS J0216-0435 is a
type 2 QSO hosted by an ERO at z=1.985. In particular, at the time of writing,
XBS~J0216-0435 is the only type~2 QSO at z$\sim$2 found in the XMM--BSS survey
(the identification rate is  $\sim$90\%). This  implies that this kind of
sources are quite rare at bright X--ray flux ($\sim$10$^{-13}$ erg s$^{-1}$
cm$^{-2}$).

\begin{figure}[t!]
{\centerline{\epsfig{file=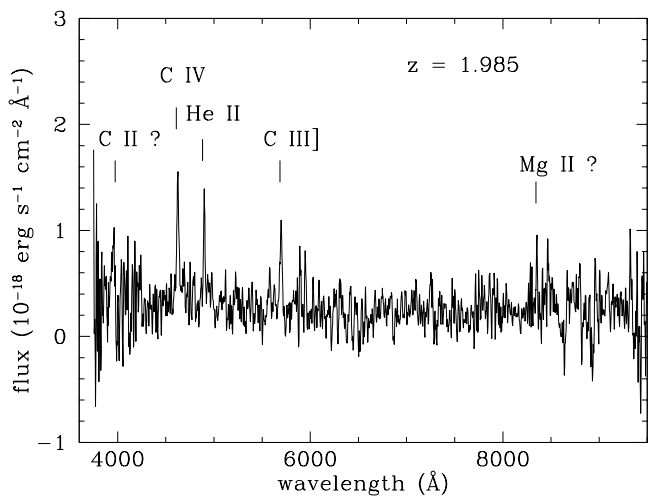, angle=-90, width=0.5\textwidth}}}
\caption{Low resolution FORS2--VLT optical spectrum of XBS J0216-0435. 
Data are not corrected for the slit loss.}
\label{fig3}
\end{figure}

\section{Spectral Energy Distribution of XBS J0216-0435}

Although the blue spectral coverage of the optical spectrum allows to give only
a weak constrain on the absorption in the optical band (A$_V$), a more
stringent lower limit can be derived by using the spectral energy distribution
(SED) of the source. All the available photometric points for XBS J0216-0435
are shown in Fig.~6. While the X--ray, 
optical and NIR data have been derived from our own analysis (see Sect. 2 and 3.1), the mid-infrared
(3.6, 4.5, 5.8, 8.0 and 24 $\mu$m observed--frame) and the radio  data (1.4 GHz
observed--frame) are taken from the literature. In particular, the
mid--infrared IRAC (the Spitzer Infrared Array Camera; Fazio et al.
\cite{fazio04} -- 3.6, 4.5, 5.8, 8 $\mu$m) and MIPS (Multi-band Imager for
Spitzer; Rieke et al. \cite{rieke04} -- 24 $\mu$m) data are part of the second
SWIRE--Spitzer data released catalogue\footnote{See
http://swire.ipac.caltech.edu/swire/public/news$\_$archive.html}  for the
XMM-LSS fields. As for the radio data, the object is covered both by the NVSS
(NRAO VLA Sky Survey, Condon et al. \cite{condon}) and by the FIRST (Faint
Images of the Radio Sky at Twenty-cm, Becker et al. \cite{becker}) surveys. In
Fig.~6 the 5$\sigma$ upper limit on the 1.4 GHz flux density derived  from the
FIRST survey, that is deeper than the NVSS, is plotted.

We compare the photometry of XBS J0216-0435 with the SED of local type 1
radio--loud (RL) and radio--quiet (RQ) QSOs adapted from Elvis et al.
\cite{elvis94}.  The two local SEDs have been normalized to the hard X--ray
data of XBS J0216-0435. As it is shown from the figure, the SED of
XBS J0216-0435 is consistent with that of a RQ--QSO. The optical/NIR data are
inconsistent with the the SED of type 1 QSO because the optical/NIR
photometric points are strongly affected by absorption.
By comparing the rest--frame
optical flux value measured for XBS J0216-0435 with the corresponding value
expected from a RQ unabsorbed AGN we can estimate a more
stringent lower limit on the optical absorption. In particular, we find that
the intrinsic dust--obscuration associated to our source is A$_V>$3 mag,
typical of type 2 AGN (e.g. Gilli et al. \cite{gilli01}). 

The mid--infrared emission of AGNs is due to the presence of dust tori
absorbing and re-radiating a significant fraction of the total luminosity of
the primary source. As it is shown from Figure 6, we find
that the near and mid--infrared emission  of XBS J0216-0435 is consistent with
that of a type 1 QSO. This implies that, if the infrared emission associated
to the AGN is not contaminated by other infrared sources (i.e. starlight
and/or galactic dust), we don't find evidence of a dependence of the
re-radiated infrared continuum with the orientation along the line of sight of
the torus (see e.g.  Pier \& Krolik \cite{pier92}, Granato et al.
\cite{granato97}, Efstathiou et al. \cite{efstathiou95}, Nenkova et al.
\cite{nenkova02}). In this case, the  X--ray/near and mid--infrared
flux ratio observed in our source is similar to those of type 1 QSO,
extending to high--z similar results obtained for high and low luminosity
AGN at lower redshit (Lutz et al \cite{lutz04}, Sturm et al. \cite{sturm06}).

\begin{figure}[t!]
{\centerline{\epsfig{file=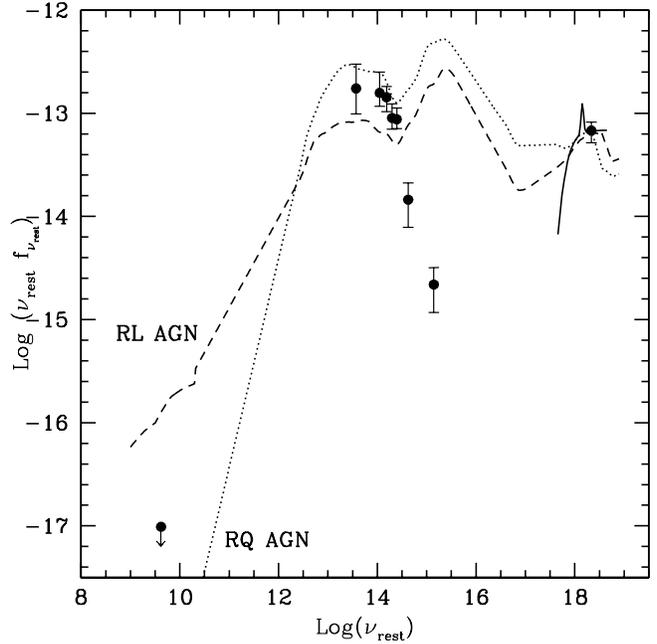, angle=0, width=0.5\textwidth}}}
\caption{Spectral energy distribution of XBS J0216-0435.
The continuous line is the X--ray best--fit model assuming the
PN normalization. Filled symbols are the  photometric points in the 
X--ray, optical, NIR, MID--IR and radio bands (see Sect.~5 for details).
The SED of local type 1 radio--loud (dashed line) and radio--quiet 
(dotted line) QSO
(adapted from Elvis et al. \cite{elvis94}) are also shown and 
normalized to the hard X--ray
data of XBS J0216-0435.}
\label{fig3}
\end{figure}

\section{QSO activity and massive galaxies}

The extended appearance in the optical and NIR bands along with the
extremely high X--ray to optical and NIR flux ratios suggest that the host
galaxy dominates the emission observed for XBS J0216-0435 at these
wavelengths.  Taking into account the redshift of the source (z=1.985)  and
given its apparent magnitude (K'$\sim$19.5 mag), the resulting rest-frame
(k-corrected) K-band absolute magnitude is M$_K$$\sim$--25.5 mag, i.e. 
$\sim$1.6 times brighter than L$^*_K$ galaxies at the same z (see Pozzetti
et al. \cite{pozzetti}; Saracco et al. \cite{saracco06}). Considering 
L*$\sim$2$\times$10$^{11}$ L$_\odot$ for galaxies at z$\sim$2 and assuming
a stellar mass--to--light ratio of 0.5 (M/L)$\odot$, we estimated a stellar
mass of the order of 10$^{11}$ $M_\odot$ for the host galaxy.

This result implies that the powerful QSO2 in XBS J0216-0435 is hosted in a
massive system. This source, along with similar results recently obtained
for other high--z type~2 QSO (see e.g. Akyiama  \cite{akiyama05},
Severgnini et al. \cite{severgnini05a} and Maiolino et al.
\cite{maiolino06}), represents  a strong observational evidence of the link
between high-z massive galaxies and powerful obscured  AGN. The X--ray and
optical spectral analysis of a bigger sample of this kind of sources will
represent one of the most efficient methods to test  a possible
co-evolution between massive galaxy and AGN activity  (Granato et al.
\cite{granato04}; Di Matteo et al. \cite{dimatteo05}; Hopkins et al.
\cite{hopkins05}) and to investigate if and how the AGN feedback can affect
the galaxies evolution (see Bower et al. \cite{bower05}, Stevens et al.
\cite{stevens05})

\section{Summary and Conclusions}

We have presented here the  X--ray and optical spectral analysis for one XMM-BSS ERO
(XBS J0216-0435) having an extremely high X--ray to optical/NIR ratio. This object is
the XMM-BSS source with the  highest F$_{(2-10~keV)}$/F$_{opt}$ ($\sim$220).  The
optical, NIR and X--ray photometric properties of  XBS J0216-0435 suggest the
presence of an X--ray obscured, optically type 2 QSO.   

From the X--ray point of view, the spectral
analysis confirms the presence of an obscured 
AGN. A statistically significant ($\sim$99\%)  excess around
2~keV in the observed--frame suggests the presence of an emission
line. By assuming that this feature corresponds to the iron K$\alpha$ line at
6.4~keV the source is at z$\sim$2 with an  intrinsic luminosity of the order
of 10$^{45}$ erg s$^{-1}$ and an intrinsic column density of $>$10$^{22}$
cm$^{-2}$, i.e. it is an X--ray obscured QSO.

Dedicated VLT spectroscopic data has allowed us to confirm the 
spectroscopic redshift (z$_o$=1.985) derived from the X--ray data.
The optical spectral analysis confirms that XBS J0216-0435 is an absorbed AGN.
In conclusion we are dealing with a high redshift, X--ray
obscured, optically type~2 QSO.

This source represents one of the few X--ray emitting EROs for which the
presence of an X-ray and optical obscured QSO can be unambiguously established
(see also Severgnini et al. \cite{severgnini05a} and Maiolino et al.
\cite{maiolino06} and references therein). Even more, XBS J0216-0435 can be
considered  one of the few examples of X--ray obscured QSO2 at high redshift
for which a detailed X--ray and optical analysis has been possible  and for
which a SED from radio to X--rays has been published (see also Sturm et al.
\cite{sturm06}).

We want to note that the physical properties derived for XBS J0216-0435 are
also in excellent agreement with the relation found by Fiore et al.
(\cite{fiore}) for type~2 AGNs. These authors show that there is a correlation 
between F$_{(2-10~keV)}$/F$_{opt}$ and L$_{2-10\rm keV}$ for optically
obscured AGNs. In particular, they find that higher luminosity AGN  have
higher  F$_{(2-10~keV)}$/F$_{opt}$.  Following this relation the
F$_X$/F$_{opt}$ of XBS J0216-0435 is consistent with a 2-10 keV luminosity of
the order of 10$^{45}$ erg s$^{-1}$ in good agreement with what has been found
in our analysis.

From the NIR data we have estimated the luminosity and the
stellar mass of the host galaxy. An example of the
coexistence between powerful QSO and massive galaxy has been found. This
result confirms that the selection of  X--ray
emitting EROs with very high X--ray--to--optical or to--NIR ratio
is a very efficient method to find not only high--z QSO but also very massive
galaxies and thus to study the fundamental connection between accretion and
star--formation processes in the Universe.

Finally, it is worth noting that XBS J0216-0435 is an X--ray line--emitting
source like those searched by  Maccacaro et al. (\cite{maccacaro04}) and 
Braito et al. (\cite{braito}) with the FLEX (Finder of Line--Emitting X--ray
sources) algorithm in the XLEO (X--ray Line--Emitting Object) project.
Therefore, a systematic search for XLEO is likely to increase the number of
know type 2 QSOs.

%______________________________________________________________

\begin{acknowledgements}

P. S. acknowledges a research fellowship from the Istituto Nazionale di
Astrofisica (INAF). This work has received partial financial support from 
ASI (ASI/INAF n. I/023/05/0), by the
Italian Ministry of  Instruction, of  University and of Research
(MIUR) through grant Cofin-03-02-23. F. J. C. and M. T. C. acknowledge
financial support from the Spanish Ministerio de Eduacion y Ciencia under
project AYA2003-00812. 
We thank C. Vignali, F. Cocchia, N. A. Webb 
for a careful reading 
of the paper and for useful comments which have improved the paper.
The TNG telescope is operated
on the island of La Palma by the Centro Galileo Galilei of the INAF in the
Spanish Observatorio del Roque de Los Muchachos of the Instituto de Astrof\'\i
sica de Canarias. We would like to thank the staff members of the ESO and TNG
Telescopes for their support during the observations . 

 \end{acknowledgements}


\begin{thebibliography}{}

  \bibitem[2001]{akiyama01}
  Akiyama, M., Ohta, K. 2001, PASJ 53, 63
  
  \bibitem[2005]{akiyama05}
   Akiyama, M. 2005, ApJ 629, 72
   
  \bibitem[2002]{alexander02}
  Alexander, D.M., Vignali, C., Bauer, F.E. et al. 2002, AJ 123, 1149
  
  \bibitem[2003]{alexander03}
  Alexander, D.M., Bauer, F.E., Brandt, W. N., et al. 2003, AJ 126, 539
   
  \bibitem[1988]{baldwin88}
   Baldwin, J. A., McMahon, R, Hazard, C., Williams, R. E. 1988, 
   ApJ 327, 103
  
  \bibitem[2003]{barger03}
  Barger, A.J., Cowie, L.L., Capak, P., et al. 2003, AJ 126, 632
  
  \bibitem[1995]{becker}
  Becker, R. H., White, R. L.,Helfand, D. J. 1995, ApJ, 450, 559
  
  \bibitem[2005]{bower05}
  Bower, R.G., Benson, A.J., Malbon, R., et al. 2005, MNRAS submitted
  [astro-ph/0511338]
  
  \bibitem[2005]{braito}
  Braito, V., Maccacaro, T., Caccianiga, A., Severgnini, P., Della Ceca, R. 
  2005, ApJ Letters, 621, 97
  
  \bibitem[2005]{brusa05} 
  Brusa, M., Comastri, A., Daddi, E., et al. 2005, A\&A 432, 69

  \bibitem[2004]{caccianiga} 
  Caccianiga, A., Severgnini, P., Braito, V., et al.  2004 A\&A 416, 910
  
  \bibitem[1998]{condon}
  Condon, J. J., Cotton, W. D., Greisen, E. W., et al. 1998, AJ, 115, 1693
  
  \bibitem[2004]{dellaceca04} 
  Della Ceca, R., Maccacaro, T., Caccianiga, A., et al. 2004, A\&A 
  428, 383
  
  \bibitem[2005]{dellaceca05} 
  Della Ceca, R., Caccianiga, A., Severgnini, P., et al. 2005
  Proc. of "The X-ray Universe 2005", San Lorenzo 
  de El Escorial (Spain) organized by the European
  Space Astronomy Centre (ESAC) of ESA
 
  \bibitem[1990]{dickey} 
   Dickey, J. M., Lockman, F. J. 1990, ARA\&A, 28, 215
   
   \bibitem[2005]{dimatteo05}
   Di Matteo, T., Springel, V., Hernquist, L. 2005, Nature, 433, 604

   \bibitem[1995]{efstathiou95}
   Efstathiou, A., Rowan-Robinson, M. 1995, MNRAS, 273, 649,
   
  \bibitem[1994]{elvis94}
   Elvis, M., Wilkes, B. J., McDowell, J. C., et al. 1994, ApJS, 95, 1

  \bibitem[2004]{fazio04}
  Fazio, G. G., Hora, J. L., Allen, L. E., et al. 2004, ApJS, 154, 10
  
  \bibitem[2003]{fiore} 
  Fiore, F., Brusa, M., Cocchia, F., et al. 2003, A\&A 409, 79  
  
  \bibitem[1995]{fukugita}
  Fukugita, M., Shimasaku, K., Icikawa, T. 1995, PASP 107, 945

  \bibitem[2004]{gandhi} 
  Gandhi, P., Crawford, C.S., Fabian, A.C., \& Johnstone, R.M. 
  2004, MNRAS, 348, 529
   
  \bibitem[2001a]{gilli}
  Gilli, R., Salvati, M., Hasinger, G., et al. 2001a, A\&A, 366, 407
  
  \bibitem[2001b]{gilli01}
  Gilli, R., Risaliti, G., Severgnini, P., et al. 2001b, ASPC, 234, 459
  
  \bibitem[1997]{granato97}
  Granato, G. L., Danese, L.,  Franceschini, A. 1997, ApJ, 486, 147
  
  \bibitem[2004]{granato04}
  Granato, G.L., De Zotti, G., Silva,,L., Bressan, A, Danese, L. 2004, ApJ
  600, 580
  
  \bibitem[1995]{heckman95}
  Heckman, T, Krolik, J., Meurer, G., et al. 1995, ApJ 452, 549
  
  \bibitem[2005]{hopkins05}
   Hopkins, P.F., Hernquist, L., Cox, T.J., et al. 2005, ApJS in press
   [astro-ph/0506398] 
  
  \bibitem[1992]{kaa}
  Kaastra J. S. 1992, An X-Ray Spectral Code for Optically Thin Plasmas.
  Internal SRON-Leiden Report, updated version 2.0

  \bibitem[1995]{lie}
  Liedahl D. A., Osterheld A. L., Goldstein W. H. 1995, ApJ Letter, 438, 115
  
  \bibitem[2004]{lutz04}
  Lutz, D., Maiolino, R., Spoon, H. W. W., Moorwood, A. F. M. 2004, A\&A, 
  418, 465

  \bibitem[1988]{maccacaro88}
  Maccacaro, T., Gioia, I. M., Wolter, A., Zamorani, G., Stocke, J.T. 
  1988, ApJ 326, 680
  
  \bibitem[2004]{maccacaro04}
  Maccacaro, T., Braito, V., Della Ceca, R., Severgnini, P., Caccianiga, A.
  2004, ApJ Letters, 617, 33
  
  \bibitem[2006]{maiolino06}
   Maiolino, R., Mignoli, M., Pozzetti, L. et al. 2006, A\&A 445, 457
  
  \bibitem[2002]{mainieri}
  Mainieri, V., Bergeron, J., Hasinger, G., et al. 2002, A\&A 393, 425
  
  \bibitem[1985]{mew}
   Mewe, R., Gronenschild, E. H. B. M., van den Oord, G. H. J. 1985, 
   A\&AS 62, 19
    
  \bibitem[2004]{mignoli}
  Mignoli, M., Pozzetti, L., Comastri, A., et al. 2004, A\&A  418, 827
  
  \bibitem[1999]{mohr}
  Mohr, J.J., Mathiesen, B., Evrard, A.E. 1999, ApJ, 517, 627.

  \bibitem[1998]{monet}
  Monet, D., Bird, A., Canzian, B., et al. 1998, The PMM USNO A2.0 Catalog 
  (Washington DC: US Naval Observatory)
  
  \bibitem[2002]{nenkova02}
  Nenkova, M., Ivezic, Z., Elitzur, M. 2002, ApJ, 570, L9
  
  \bibitem[1992]{pier92}
  Pier, E.A., Krolik, J.H. 1992, ApJ, 401, 99
  
  \bibitem[2003]{pozzetti}
  Pozzetti, L., Cimatti, A., Zamorani, G., et al. 2003, A\&A 402, 837
  
  \bibitem[1994]{reynolds}
  Reynolds, C.S., Fabian, A.,C., Makishima, K., Fukazawa, Y., Tamura, T.
  1994, MNRAS 268, L55.
  
  \bibitem[2004]{rieke04}
  Rieke, G. H., Young, E. T., Engelbracht, C. W., et al. 2004, ApJS, 154, 25
  
  \bibitem[2003]{roche}
  Roche, N.D., Dunlop, J., Almaini, O. 2003, MNRAS 346, 803
  
  \bibitem[2006]{saracco06}
  Saracco, P., Fiano, A., Chincarini, G., et al. 2006, MNRAS in press
  [astro-ph/0512147]

  \bibitem[2003]{severgnini03} 
  Severgnini, P., Caccianiga, A., Braito, V., et al. 2003, A\&A 406, 483
  
  \bibitem[2005a]{severgnini05a} 
  Severgnini, P., Della Ceca, R., Braito, V., et al., 2005a, A\&A
  431, 87
  
  \bibitem[2005b]{severgnini05b}
  Severgnini, P., Braito, V., Caccianiga, A., et al. 2005b, 
  Proc. of "The X-ray Universe 2005", San Lorenzo 
  de El Escorial (Spain) organized by the European
  Space Astronomy Centre (ESAC) of ESA
  
  \bibitem[2005]{stevens05}
  Stevens J.A., Page M.J., Ivison R.J., et al. 2005, MNRAS, 360, 610

  \bibitem[2006]{sturm06}
  Sturm, E., Hasinger, G., Lehmann, I., et al. 2006, ApJ in press
  
  \bibitem[2004]{szokoly}
  Szokoly, G.P., Bergeron, J., Hasinger, G., et al. 2004, ApJSS 155, 271
  
  \bibitem[2003]{stevens}
  Stevens, J.A., Page M.J., Ivison R.J., et al. 2003 MNRAS 342, 249
  
  \bibitem[2003]{ueda}
  Ueda, Y., Akiyama, M., Ohta, K., Miyaji, T. 2003, ApJ, 598, 886
     
  \bibitem[2003]{willott}
  Willott, C.J., Simpson, C., Alamini, O., et al. 2003, MNRAS 339, 397

  \end{thebibliography}
\end{document}